# Incremental capacity-based multi-feature fusion model for predicting state-of-health of lithium-ion batteries

**Chenyu Jia, Guangshu Xia, Yuanhao Shi, Jianfang Jia, Jie Wen[*], Jianchao Zeng**

School of Electrical and Control Engineering, North University of China, Taiyuan 030051, China

**Abstract**

Lithium-ion batteries have become an indispensable part of human industrial production and daily life. For the safe use, management and maintenance of lithium-ion batteries, the state of health (SOH) of lithium-ion batteries is an important indicator so that the SOH estimation is of significant practical value. In order to accurately predict SOH, this paper proposes a fusion prediction model which combines particle swarm optimization (PSO) algorithm, bi-directional long-short time memory network (BiLSTM) and adaptive boosting (AdaBoost) algorithm. In the proposed prediction model, indirect health indicators (HIs), which characterize battery degradation, are obtained with the help of incremental capacity analysis (ICA), and is fed into BiLSTM to extract time-series features, whose parameters are optimized by employing PSO algorithm. On this basis, the AdaBoost algorithm is applied to reduce the risk of overfitting the PSO-BiLSTM model. The study based on lithium-ion battery data from Center for Advanced Life Cycle Engineering (CALCE) shows that the PSO-BiLSTM-AdaBoost model has higher accuracy, better robustness, and generalization ability.

## 1. Introduction

Since the 21st century, with the increasing energy insufficiency and environmental degradation, the demand for new energy is increasing day by day, and the scale of the new energy field is also rising [1]. Against this background, benefiting from the high energy density, small volume and low cost, lithium-ion batteries (LiBs) have been widely used in electronic mobile devices, EVs and grid energy storage systems [2, 3]. However, due to the characteristics of the LiBs themselves, LiBs will inevitably age during use [4], so that they need to be replaced at the first opportunity in order to avoid causing property damage or even casualties. Therefore, the accurate prediction of SOH of LiBs is crucial in prognostics and health management (PHM) and battery management systems (BMS) [5].

However, since the degradation trend of LiBs is non-linear and different use states may lead to different degradation patterns, accurately estimating the health state of a LiB is a very challenging task. According to the review articles [6, 7], the current stage of study addressing the degradation prediction of LiBs can be

[*] Corresponding author: Jie Wen (wenjie015@gmail.com)

This work was supported in part by National Natural Science Foundation of China under Grant 72071183 and in part by the Natural Science Foundation of Shanxi Province under Grant 202403021211088.

broadly categorized into three schools of thought: model-based methods, data-driven methods and fusion methods. Model-based methods are classified into electrochemical models [8-10], equivalent circuit models [11-14] and empirical models [15-18]. Although model-based methods are able to reflect the degradation behavior and chemical properties of batteries in different operating states, modelling requires a full understanding of the reaction mechanisms within LiBs, which adds to the difficulty of building accurate models.

Unlike model-based methods, data-driven methods, which are mainly based on machine learning, are more concerned with mining useful information from historical degradation data of batteries and do not rely on a priori knowledge. In addition, the time spent on repetitive modeling can be greatly reduced due to the good portability of the data-driven approach. Data-driven methods can be categorized into direct prediction and indirect prediction [19]. In direct prediction, the capacity and impedance of the battery are used as inputs to the prediction, but they are difficult to obtain and the measuring instruments are very expensive. Indirect prediction refers to the use of data such as voltage, current and temperature from battery operation as features, i.e., HI. Classical data-driven methods include relevance vector machine (RVM) [20], support vector machine (SVM) [21, 22], support vector regression (SVR) [23-26] and artificial neural networks (ANN) [27-29]. For example, Yang et al. fused two- and three-dimensional convolutional neural networks (CNN) to improve the prediction performance by feature-focused algorithms and multi-scale cycle-focused algorithms [30], while Shen et al. proposed a deep CNN combined with integrated learning strategy that was tested on the implantable LiB dataset [31]. Since the Gaussian regression process (GPR) can provide probabilistic predictions, Pan et al. constructed a GPR based on incremental capacity curves and obtained more satisfactory results [32]. Recurrent Neural Networks (RNN) had also been used to predict the degradation of battery performance due to its excellent performance [33], and some variants of RNNs are widely used in this field. Luo et al. proposed a method based on hybrid data preprocessing and bi-directional long-short term memory network (BiLSTM) combined with a hierarchical attention mechanism, which was validated on the spacecraft LiB dataset and effectively improved the prediction performance [34]. In recent years, incremental capacity analysis (ICA) and differential thermal capacity analysis (DTC) have become research hotspots in the field of SOH prediction due to their ability to respond to a certain extent to the internal changes of LiBs. For instance, by analyzing the IC curve, Pang et al. selected the peak of the IC curve and the area under the curve for characterizing the degradation process of LiBs [35]. Tang et al. reconstructed the IC curve based on the current variation curve, and the SOH of the cell can be quickly determined by a simple mapping [36]. The ageing of commercial LiFePO4 batteries has been explored according to ICA, differential voltage analysis (DVA) as well as open-circuit voltage (OCV), and the incremental capacity curves they obtained can be used to predict the ageing of the batteries [37]. The team led by Lin extracted multiple features from degradation information such as IC and fused the SOH results using a random forest (RF) model [38]. Feng et al. de-fitted the established functional model based on DTC and used an extreme learning machine for SOH prediction [39]. Zhao et al. used fuzzy information granularity to shorten the sequence of LiBs, giving predictions for the upper and lower bounds, respectively, to achieve interval-like predictions [40].

Although there have been many studies that have taken the ICA approach, however, more often than not, only some kind of single information in it, such as area under the curve and peaks, is obtained as a kind of

HI. Although more accurate results are eventually obtained, other information contained in the IC curves is ignored. Moreover, most of the data-driven methods use only one machine learning method and tend to follow empiricism in parameter setting, which may result in that the constructed models are not in an optimal state.

In order to solve the problems of insufficient information extraction for IC curves and the parameterization of machine learning models that often follow experience in previous studies, a multi-feature fusion model based on IC curves is designed in this paper to predict the SOH of LiBs. Multiple features are extracted from IC curves to describe the degradation behavior of LiBs more comprehensively. Considering the capacity recovery of LiBs during degradation [41], we use BiLSTM model to predict SOH and optimize the model parameters by PSO algorithm. In order to avoid the prediction results of the model falling into local optimum, inspired by the advantages of the convergence model, the AdaBoost algorithm is chosen in this paper to be used for constructing the fusion prediction model. The experiments based on CALCE dataset show that the proposed PSO-BiLSTM-AdaBoost model exhibits higher accuracy in the prediction of SOH for LiBs. Therefore, the main contributions of this paper can be summarized in the following four points. (1) Multiple indirect HIs are extracted from IC curves, replacing the traditional HIs for SOH prediction, and data redundancy is prevented by a dimensionality reduction algorithm. (2) The PSO-BiLSTM model, which can better capture the information contained in future sequences and past sequences than the widely used unidirectional LSTM model. (3) By combining BiLSTM and AdaBoost algorithm, a PSO-BiLSTM-AdaBoost model is proposed to improve the SOH prediction accuracy of LiBs and prevent overfitting through an integrated learning approach. (4) Based on publicly available LiB datasets, the effectiveness and reliability of the proposed SOH prediction model are verified and evaluated by using four difference metrics.

Subsequent sections of this paper follow. Section II describes the LiB degradation dataset and the feature extraction method. Section III describes the proposed PSO-BiLSTM-AdaBoost model along with the associated SOH prediction procedure. Section IV presents the results of the SOH prediction and its qualitative and quantitative analysis. Section V summarizes the paper with future directions.

**2. Battery degradation datasets and feature extraction**

This section delves into the dataset concerning battery degradation utilized within this paper. The commonly used HI is obtained for the charge/discharge data of the battery first, and then a variety of eigenvalues, which are used as inputs to the SOH prediction model, are extracted from the IC curves and subjected to dimensionality reduction.

*2.1. Datasets and common HI extractions*

*2.1.1. Datasets*

The ageing of LiBs is usually characterized by the decay of the available capacity, which is used to define SOH

$$\text{SOH} = \frac{C_{\text{Reality}}}{C_{\text{Nominal}}} \quad (1)$$

where, $C_{\text{Reality}}$ and $C_{\text{Nominal}}$ denote the current maximum available capacity and nominal capacity, respectively.

In this study, we chose the LiB dataset released by CALCE for experimentation. Batteries CS2-35, CS2-36, CS2-37, and CS2-38 with a nominal capacity of 1.1Ah were selected for the study due to that the batteries CS2 series have a longer cycle aging period than the commonly used NASA dataset. The batteries CS2 have the same charging profile, are charged using the standard constant-current/constant-voltage (CC/CV) process, and discharged at a constant discharge rate of 1 C. Meanwhile, the batteries CS2 have the same charge curve, are charged using the standard CC/CV process, and are discharged at a constant discharge rate of 1.0 C. More information on the dataset is summarized in Table 1, and the variation of CALCE battery capacity with the number of cycles are shown in Fig. 1.

Table 1 Information of CALCE LiB Dataset

| Battery | Rated capacity(Ah) | Failure threshold(Ah) | Charge cut-off voltage(V) | Discharge cut-off voltage(V) | Discharge current(A) |
|---|---|---|---|---|---|
| CS2-35 | 1.1 | 0.77 | 4.2 | 2.7 | 1 |
| CS2-36 | 1.1 | 0.77 | 4.2 | 2.5 | 1 |
| CS2-37 | 1.1 | 0.77 | 4.2 | 2.3 | 1 |
| CS2-38 | 1.1 | 0.77 | 4.2 | 2.5 | 1 |

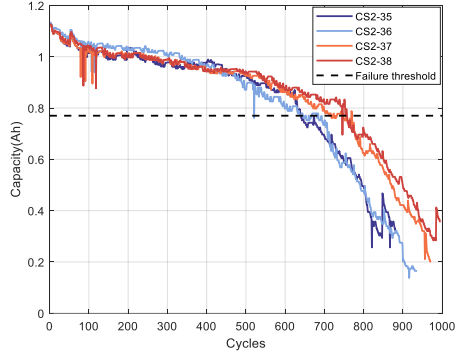

**Fig. 1.** Capacity degradation trends of LiB

From Fig. 1, we can reach an obvious conclusion: the pre-decline process of battery capacity is inherently non-linear, but shows an approximately linear trend in the overall picture. As the cycle count increases, the capacity degradation curve exhibits a non-linear nature and shows a certain degree of capacity recovery. For the data published by CALCE, we can also see from Fig. 1 that the battery capacity shows an accelerated decreasing trend after a number of cyclic aging experiments [42], and Ref. [43] suggests that the sudden drop in the capacity curve explains the lithium plating phenomenon during charging, which makes the SOH prediction of LiBs more difficult. Although the HI obtained based on voltage, current and temperature in previous studies can characterize the decline of battery capacity, it does not mechanistically reflect the internal condition of the battery. Therefore, in this paper, we perform high-precision state of health (SOH) estimation by analyzing the IC curves of the discharge phase and extracting new features from them.

*2.1.2 Common HI Extraction*

Referring to the work done by Liu et al [44], we extracted several HIs for CALCE, such as constant current charging time (CCCT), constant voltage charging time (CVCT), constant current discharge time (CCDT), and internal resistance. In order to better show the relationship between common HIs and capacity, taking the battery CS2-35 as an example, we normalized the CCCT, CVCT and CCDT and fitted them to the capacity respectively, which is shown in Fig. 2. Considering that it is more difficult to obtain the internal resistance of batteries in practice, we chose CCCT, CVCT and CCDT as the common HIs.

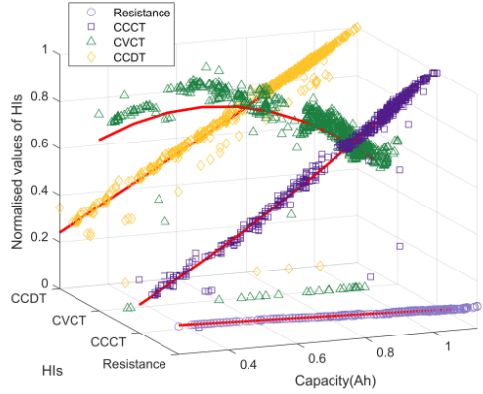

**Fig. 2.** Common HIs of CS2-35

*2.2. ICA-based HI extraction*

*2.2.1 ICA*

As mentioned in [32], the use of HIs to map the battery capacity is well able to handle the accelerated capacity degradation in terms of SOH prediction.

The ICA curve describes the change in capacity per unit voltage during charging and discharging of a battery, and in order to make the study more close to the practical application, we choose the ICA curve of the discharging stage here, whose calculation formula is

$$ICA = \frac{dQ}{dV} = \frac{dQ/dt}{dV/dt} = \frac{I \times dt}{dV} = \frac{\Delta Q}{\Delta V} \quad (2)$$

where $I, t, V$ and $Q$ represent the current, time, voltage, and discharge capacity during constant current discharge, respectively.

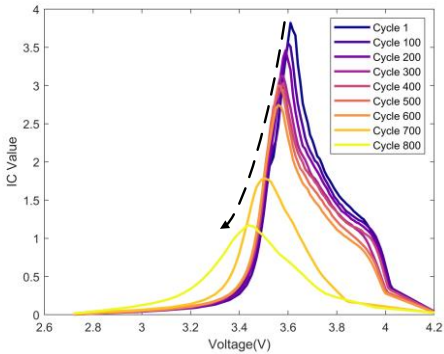

**Fig. 3.** IC curves for partial cycling of CS2-35

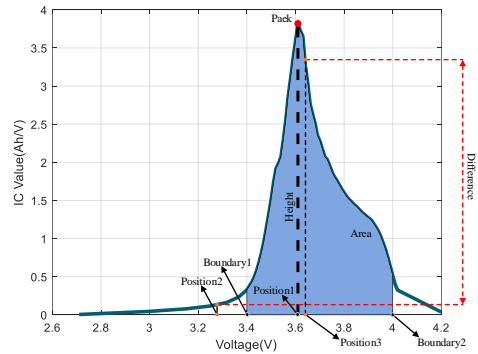

**Fig. 4.** IC curve-based feature extraction scheme

Since the IC curves are discrete in nature, it is difficult to extract the features directly, so we perform filtering operations on the extracted IC curves separately to obtain smoother curves. Inspired by [27], we use the Savitzky-Golay filtering method for the CALCE dataset. In order to reduce the possible loss of information due to filtering, we set the sliding window size to 9 and the polynomial order to 3. The smoothed ICA curves are shown in Fig. 3, which shows the discharged ICA curves of the partial cycle of the battery CS2-35 as an example. It is easy to see that the overall tendency of the curve moves towards the low voltage region as the aging experiment proceeds.

*2.2.2 HI extraction*

In previous studies, people are more likely to use the area under the IC curve or the peak value as HI, lacking other forms of information extraction. As we can see from Fig. 3, in the pre-cycle of the battery (take cycle 1-600 as an example), the IC curve shows a "convex" characteristic in the high-voltage region (4.0V-3.8V). As the battery ages (after 700 cycles), this characteristic disappears. Thus, it is important to define the area under the curve to obtain the HI of LiBs. In this paper, the voltage at which the IC curve peaks is determined first, and then the area in different voltage ranges is integrated, and the highest correlation is taken as the HI. The specific feature selection scheme is shown in Fig. 4.

A. Selection the region beneath the curve

In previous studies, the full region beneath the ICA curve is usually calculated as a HI. This is because the area under the ICA curve, which is used to characterize the change in remaining usable capacity as the LiB ages, is the difference in LiB capacity between two voltage thresholds (Boundary1 and Boundary2) as shown in the following equation

$$Area = \int_{Boundary1(V)}^{Boundary2(V)} ICA(V) dV = \int_{Boundary1(V)}^{Boundary2(V)} \frac{dQ(V)}{dv} dV = Q_{Boundary2(V)} - Q_{Boundary1(V)} \quad (3)$$

Although this approach certainly gives the full information contained in the area, it does not reflect exactly which portion of the area reflects the capacity decline.

In this paper, we identify the *Peak* of the ICA curve to obtain the corresponding voltage point $Position1(V)$, followed by integrating the area under the curve according to the defined voltage intervals $Boundary2(V)$ and $Boundary1(V)$, which is represented as Area in Fig. 4. Our range in choosing the Boundary is defined from 0.42 V to 0.38 V, and the voltage step in the range selection is 0.02 V. On this basis, we let $Boundary2(V) = Position1(V) + 0.42V$ and $Boundary1(V) = Position1(V) - 0.38V$. It can be calculated that the Pearson correlation coefficient between this HI and capacity reaches 0.9969.

B. Difference between corresponding IC values at fixed voltage

$\Delta \frac{dQ}{dV}$ has been chosen in [27] as the HI to predict SOH and RUL on NASA dataset. In this paper, we use the following technical route to obtain this HI: determine the fixed voltage points $Position2(V)$ and $Position3(V)$, then find the corresponding IC values of each and make the difference, which is manifested as the difference in Fig. 4, denoted as $A_i$ and calculated by

$$A_1 = \frac{dQ}{dV}\bigg|_{Position3=3.90V} - \frac{dQ}{dV}\bigg|_{Position2=3.40V} \tag{4}$$

$$A_2 = \frac{dQ}{dV}\bigg|_{Position3=3.80V} - \frac{dQ}{dV}\bigg|_{Position2=3.20V} \tag{5}$$

$$A_3 = \frac{dQ}{dV}\bigg|_{Position3=3.74V} - \frac{dQ}{dV}\bigg|_{Position2=3.16V} \tag{6}$$

$$A_4 = \frac{dQ}{dV}\bigg|_{Position3=3.70V} - \frac{dQ}{dV}\bigg|_{Position2=3.10V} \tag{7}$$

$$A_5 = \frac{dQ}{dV}\bigg|_{Position3=3.68V} - \frac{dQ}{dV}\bigg|_{Position2=3.12V} \tag{8}$$

$$A_6 = \frac{dQ}{dV}\bigg|_{Position3=3.52V} - \frac{dQ}{dV}\bigg|_{Position2=3.48V} \tag{9}$$

Along this line, we calculate various voltage combinations. Taking battery CS2-35 as an example, Table 2 demonstrates their correlation coefficients with capacity. Eventually, we chose $A_4$ as the HI representing $\Delta\frac{dQ}{dV}$.

Table 2 IC difference and capacity correlation coefficients for different voltage points

| HI | $A_1$ | $A_2$ | $A_3$ | $A_4$ | $A_5$ | $A_6$ |
|---|---|---|---|---|---|---|
| Pearson | 0.8892 | 0.9814 | 0.9849 | 0.9924 | 0.9841 | 0.8634 |

C. Dimensionless eigenvalue extraction based on IC curves

Dimensionless statistical eigenvalues are often used in fault diagnosis of gears for reflecting the magnitude of impact energy. Considering that each discharge cycle possesses an independent IC curve, we try to extract the dimensionless features from the IC curve. In time domain analysis, the commonly used dimensionless features are Crest Factor (CF), Pulse Factor (PF), Marginal Factor (MF), Waveform Factor (WF), and Kurtosis (Kur). For each IC curve, as a discrete sequence, its dimensionless features can be calculated by

$$CF = \frac{a_p}{a_{rms}} = \frac{\max|a_i|}{\sqrt{\frac{1}{N}\sum_{i=1}^{N}a_i^2}} \tag{10}$$

$$PF = \frac{a_p}{\bar{a}} = \frac{\max|a_i|}{\frac{1}{N}\sum_{i=1}^{N}|a_i|} \tag{11}$$

$$MF = \frac{a_p}{a_r} = \frac{\max|a_i|}{(\frac{1}{N}\sum_{i=1}^{N}\sqrt{|a_i|})^2} \tag{12}$$

$$Kur = \frac{1}{N}\sum_{i=1}^{N}\frac{a_i^4}{(\frac{1}{N}\sum_{i=1}^{N}a_i^2)^2} - 3 \tag{13}$$

In addition to the above features, we extracted IC curves difference in mean value (DMV), peak (P), maximum positive slope (MPS), peak-to-position value (PPV), average rate of change (ARC) and peak corresponding voltage (PCV), totaling 13 HIs.

*2.2.3 Correlation analysis and PCA*

In subsubsection 2.2.2, we extracted a variety of HIs based on the IC curves of the discharge stage, and their Pearson correlation heat map with the capacity is shown in Fig. 5, where the more the color is inclined to green, the stronger the correlation with the capacity; on the contrary, if the color is inclined to blue, the lower the correlation. It can be seen from Fig. 5 that the preliminary selected indicators can reflect the capacity degradation trend, and most HIs have a strong positive correlation with the cyclic aging process of the capacity. Hence, the aforementioned HIs can be utilized for predicting the SOH.

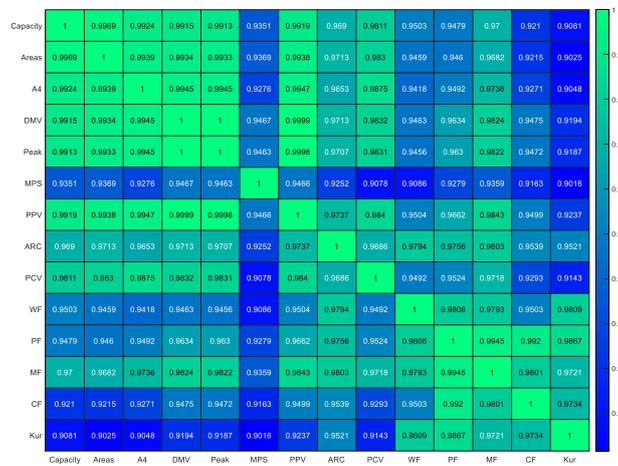

**Fig. 5.** HI vs. Capacity Correlation Heat Map for CS2-35

The key to accurately predicting SOH is to fully utilize valid information while eliminating the impact of redundant information, because the redundant information will interfere with the training process and lead to poorer prediction. Therefore, based on the previous analysis, MPS, PF, CF and Kur are eliminated first, and then in order to further carry out the elimination of redundant features, we apply principal component analysis to the remaining nine HIs for dimensionality reduction.

In subsection 2.1.2, the three HIs commonly used in CALCE are extracted, and ensure that the input data are dimensionally consistent, we chose to reduce the nine types of feature data to three dimensions. Taking the calculation results of battery CS2-35 as an example, as shown in Table 3, the eigenvalues and contribution rates of each principal component are obtained, and the contribution rate of the principal component with the serial number of 1 reaches more than 98%.

Table 3 Principal components and their contributions

| Serial number | Eigenvalue | Contribution rate /% | Cumulative contribution rate |
| --- | --- | --- | --- |
| 1 | 3.5376 | 98.97 | 98.97 |
| 2 | 0.0328 | 0.92 | 99.89 |
| 3 | 0.0031 | 0.09 | 99.98 |
| 4 | 0.0007 | 0.00 | 100.00 |
| 5 | 0.0001 | 0.00 | 100.00 |
| 6 | 0 | 0.00 | 100.00 |
| 7 | 0 | 0.00 | 100.00 |
| 8 | 0 | 0.00 | 100.00 |
| 9 | 0 | 0.00 | 100.00 |

## 3. Methodology

In this section, we construct the PSO-BiLSTM-AdaBoost model to predict the SOH of LiBs. BiLSTM is used to predict SOH, while PSO optimization algorithm is to find the number of hidden layer units and learning rate of BiLSTM. In order to prevent the overfitting problem and the local optimum problem that the PSO algorithm may fall into, the AdaBoost method is introduced to integrate PSO-BiLSTM, construct a certain number of weak predictors, and finally form a strong predictor to get the final SOH prediction.

*3.1. LSTM and BiLSTM*

In 1986, David Rumelhart innovatively introduced the concept of RNN [45], which emphasizes more on the correlation between data than traditional neural networks and determines the output of the current state based on past memories. In order to solve the problem of gradient vanishing or gradient explosion that exists in RNNs [41], the LSTM network structure was proposed [46]. A standard LSTM unit is shown in Fig. 6, which has three inputs and two outputs.

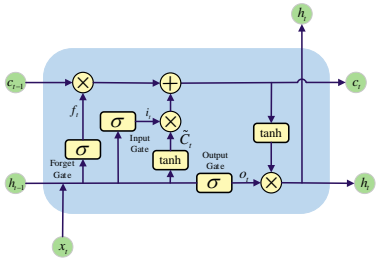
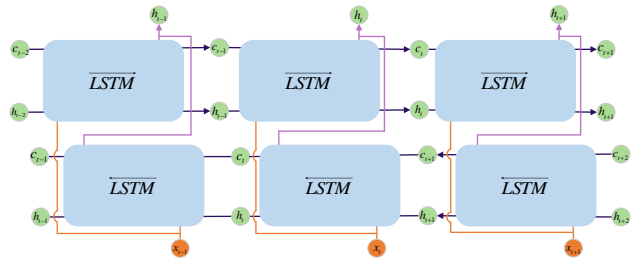

**Fig. 6.** Standard LSTM unit         **Fig. 7.** BiLSTM network structure

The distinctive characteristic of LSTM lies in the primary thread running through its base unit, which records the unit state of the neuron and covers the neural network's generalization and summarization of all the input information before the next moment. LSTM has three gating controls: forgetting gate, input gate, and output gate, and for the unit of the current moment it first determines whether or not the information $h_{t-1}$ in the inputs is retained. The $W$ and $b$ in the following equations represent the weight and bias of the corresponding gating respectively. The forgetting gate can be described by

$$f_t = \sigma\left(W_f \cdot [h_{t-1}, x_t] + b_f\right) \tag{14}$$

where, $x_t$ represents the input value and $\sigma$ represents the sigmod activation function.

The role of the input gate is to update the state of the cell, determine the weight value of the current information being input into the cell, and then go through the tanh layer to get the value of the candidate storage cell $\tilde{C}_t$, which is given as

$$i_t = \sigma\left(W_i \cdot [h_{t-1}, x_t] + b_i\right) \tag{15}$$

$$\tilde{C}_t = \tanh\left(W_c \cdot [h_{t-1}, x_t] + b_c\right) \tag{16}$$

Combining the outputs of the forgetting gate and the input gate, the cell state $C_t$ is updated by where $C_{t-1}$ represents the state value of the previous LSTM cell, i.e.

$$C_t = f_t * C_{t-1} + i_t * \tilde{C}_t \tag{17}$$

Finally, the output gate and tanh layer generate the output of the network $h_t$, describing by

$$o_t = \sigma\left(W_o \cdot [h_{t-1}, x_t] + b_o\right) \tag{18}$$

$$h_t = o_t * \tanh(C_t) \tag{19}$$

Although LSTM can memorize previous information, it ignores the effect of future information on current information. Thus, the BiLSTM model is a better choice when dealing with long sequence problems, which can refer to both the information before and after the moment to be predicted. The architecture of BiLSTM is illustrated in Fig. 7.

BiLSTM consists of two layers of LSTMs propagated in the forward and backward directions, and the final prediction for each time step is a merger of the results from both directions. BiLSTM can be defined by

$$\begin{cases} h_t^+ = \overrightarrow{LSTM}\left(h_{t-1}, W_t, C_{t-1}\right) \\ h_t^- = \overleftarrow{LSTM}\left(h_{t+1}, W_t, C_{t+1}\right) \\ h_t = \left[h_t^+, h_t^-\right] \end{cases} \tag{20}$$

and the hyperparameters of the BiLSTM model in this paper are shown in Table 4.

Table 4 Parameter settings

| Parameters | Value |
| --- | --- |
| Optimizers | Adam |
| Loss function | MSE |
| Activation function | RELU |
| LearnRateDropPeriod | 350 |
| LearnRateDropFactor | 0.01 |
| Min Batch size | 32 |
| Max Epoch | 500 |

*3.2. Evaluation indicators*

In this paper, four error assessment metrics are used to quantitatively assess the output of the model, namely Root Mean Square Error (RMSE), Mean Absolute Error (MAE), Mean Absolute Percentage Error (MAPE), and Mean Square Error (MSE), which are defined as

$$RMSE = \sqrt{\frac{1}{N}\sum_{i=1}^{N}(x_i - \hat{x}_i)^2} \tag{21}$$

$$MAE = \frac{1}{N}\sum_{i=1}^{N}|x_i - \hat{x}_i| \tag{22}$$

$$MAPE = \frac{1}{N}\sum_{i=1}^{N}\left|\frac{x_i - \hat{x}_i}{x_i}\right| \times 100\% \tag{23}$$

$$MSE = \frac{1}{N}\sum_{i=1}^{N}(x_i - \hat{x}_i)^2 \tag{24}$$

where, $x_i$ denotes the true value of SOH and $\hat{x}_i$ denotes the predicted value of SOH. The smaller their values, the higher the prediction accuracy.

*3.3. PSO and AdaBoost*

A. PSO

Particle Swarm Optimization (PSO) algorithm performs a global stochastic search through the generated particles to find the best solution to the objective. PSO has been widely used in model optimization due to its ease of implementation and fast convergence.

In some of the existing studies, the number of units of the neural network is usually set to a fixed value, which causes the generalization performance of the neural network to deteriorate. In this paper, we use the PSO algorithm to find two parameters of the BiLSTM model: the number of units in the hidden layer and the learning rate. The population number of PSO is set to 20, the learning rates $c_1$ and $c_2$ are set to 1.2 and 1.8, respectively, and the inertia weights are in the range of [0.2,1.1].

B. AdaBoost

In order to avoid falling into a local optimum as the number of PSO iterations increases, we further optimize the model using the AdaBoost algorithm, which is an integrated learning method firstly applied to classification problems, and which combines multiple weak predictors with different weights to form a strong predictor. The specific steps are as follows: (1) train the first weak predictor using the initialized sample weights; (2) update the weights according to the error to obtain different weak predictors; (3) combine all the weak predictors in a weighted way to obtain strong predictions. In order to avoid poor results due to too few predictors and expensive time costs due to too many predictors, this paper chooses to determine the number of weak predictors as 10.

*3.4. Forecasting process*

From the architecture of PSO-BiLSTM-AdaBoost model, the flowchart depicting SOH prediction is presented in Fig. 8, and the process of SOH prediction requires four steps.

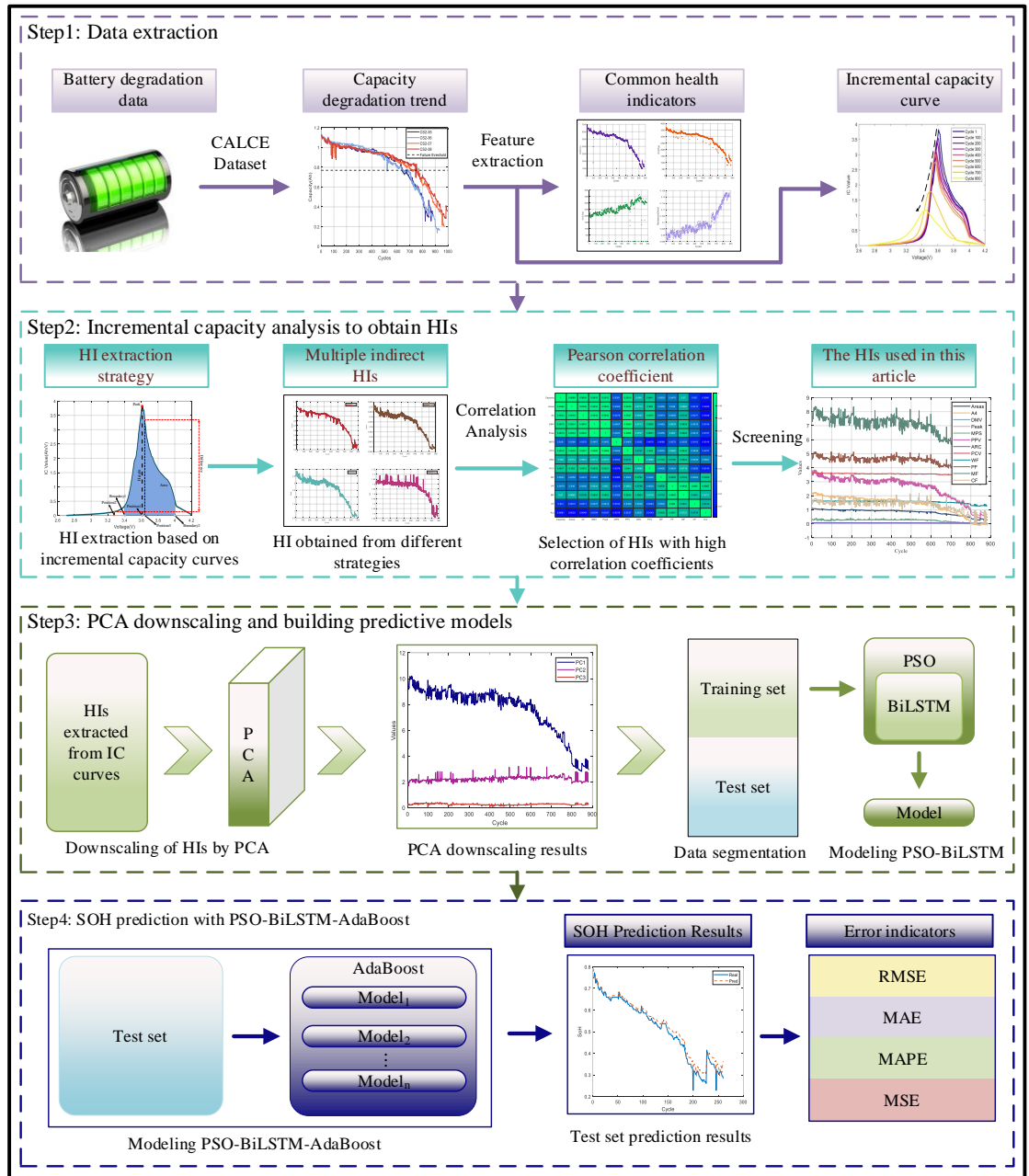

**Fig. 8.** SOH Prediction flowchart of PSO-BiLSTM-AdaBoost model.

1) Data extraction.

By processing the raw sampling data, we obtained the capacity decay curve of LiBs cyclic aging and four commonly used HIs based on the data such as charging and discharging voltages. On this basis, the incremental capacity curve of LiBs for the discharging process is obtained.

2) Incremental capacity analysis to obtain HIs.

Based on the incremental capacity curves extracted in the previous step, we obtained a variety of HIs from them reflecting the aging of the batteries. Through correlation analysis, we verified the credibility and validity of these HIs characterizing the capacity of the batteries, and excluded some of the poorly behaved HIs in order to prevent the data redundancy from interfering with the prediction of the SOH.

3) PCA downscaling and building predictive models.

In this step, we subject the processed HIs to principal component analysis dimensionality reduction to further reduce data redundancy first. Then, the processed dataset is divided into training set and test set. According to the general logic, the higher the share of training set, the higher the accuracy of prediction. However, since the total amount of data is fixed, if the training set is too large, it will lead to a prediction starting point close to the failure threshold, making the prediction effect not meaningful. Therefore, in the SOH prediction experiments, we select 50%, 60% and 70% of the entire dataset as the training set, respectively, after which we use the divided training set to train the PSO-BiLSTM model.

4) SOH prediction with PSO-BiLSTM-AdaBoost.

In this step, the test set is fed into the optimized training model and the output of the model is used as the predicted value of SOH. In order to verify the effectiveness of the proposed prediction model, we will experimentally evaluate the prediction performance of the PSO-BiLSTM-AdaBoost model using the error assessment metrics mentioned in subsection 3.2 and analyze its performance in prediction by comparing it with different models.

Following the flowchart, in the next section, we will use the proposed PSO-BiLSTM-AdaBoost model to predict the SOH of LiBs and compare and analyze the prediction results.

## 4. Experiment results and analysis

In this chapter, the SOH prediction experiments are conducted based on the HI in Section 2 and the model constructed in Section 3 to verify that our proposed prediction model is valid. To verify the superior performance of the prediction model, we compare it with other models.

*4.1. Comparison and Analysis of Prediction Results for Different HIs*

Based on the work in Section 2, we make the traditional HI and the feature selection method proposed in this paper as inputs for predicting SOH, respectively. As an example, we use the BiLSTM model for prediction with the number of neurons of the model set to 60 and the learning rate of 0.01. The error results obtained are shown in Fig. 9 and Fig. 10, where Fig. 9 shows the MAE, RMSE and MSE of the prediction results for different inputs, with three sets of errors from left to right, representing 50%, 60% and 70% of the training data, respectively, while the MAPE is displayed in Fig. 10. In Figs. 9 and 10, Ours denotes the feature by using our proposed feature extraction strategy and Normal HI denotes the traditional three features.

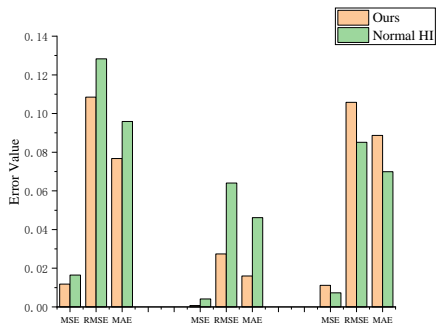 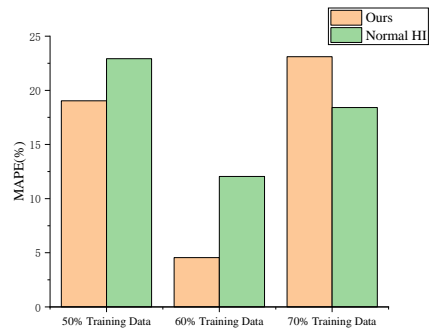

**Fig. 9.** Prediction errors for different HI  **Fig. 10.** MAPE with different training data

*4.2. SOH prediction and results analysis*

In this subsection, we use the proposed PSO-BiLSTM-AdaBoost model to predict SOH at different starting points.To extensively compare the performance of the models, we design SVM, RF, SVR, ELM-AdaBoost, CNN, PSO-BP, PSO-BiLSTM, and PSO-GRU-AdaBoost totaling 8 comparison models. Taking the battery CS2-35 as an example, when the training data is 50%, the prediction results are shown in Fig. 11, where Ours represents the proposed PSO-BiLSTM-AdaBoost model. From Fig. 11, it can be seen that the prediction results of SVM, SVR and PSO-BP are very poor, and the results obtained are far from the degradation trajectory of the battery. The LSTM model and its derived structure show superior predictive performance compared to traditional machine learning models. Especially for the data after 820 cycles, the predicted values obtained by several traditional models are strongly volatile. Obviously, they do not cope well with sudden data drops. In addition, for the last 100 cycles of the battery CS2-35, the method proposed in this paper achieves the best prediction. For BiLSTM, finding the optimal parameters by PSO and combining with AdaBoost algorithm makes the prediction results closer to the real values. In addition, in this experiment, although the PSO-BiLSTM method has the best fit to the true values for the predictions from the prediction start point to 650 cycles, it performs weaker than our proposed model in the interval from 750 to the end. This also reflects another conclusion that for a certain model, it will perform very well in a certain interval of specific data, but lacks generalizability.

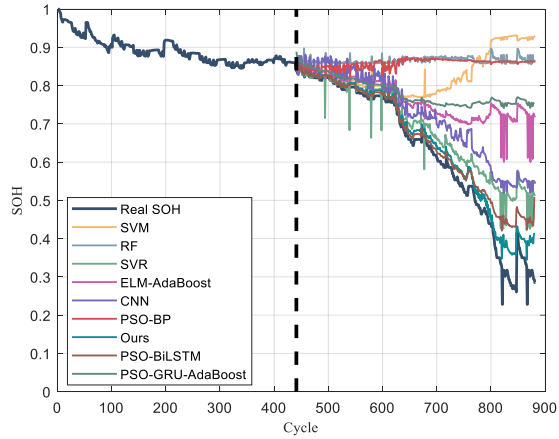

**Fig. 11.** SOH prediction for battery CS2-35 using different models at 50% of training data

To further validate the effectiveness of our method on different data, we performed multiple sets of experiments on the CALCE data. The prediction results for batteries CS2-35, CS2-36, and CS2-37 are shown in Fig. 12. For battery CS2-35, the results at 60% and 70% of the training data are shown in Fig. 13(a) and (b), from which it can be seen that although the prediction of the traditional machine learning model is significantly improved with the increase of training data, it is still weaker than our proposed method in terms of results. Based on what is presented in Fig. 12, our method shows the best prediction performance and the prediction results are the closest to the real SOH. the main reason is that BiLSTM is able to learn the input features in both directions to reduce the prediction error and closer to the real value than LSTM, and AdaBoost algorithm can improve the generalization performance of the model to avoid the risk of overfitting.

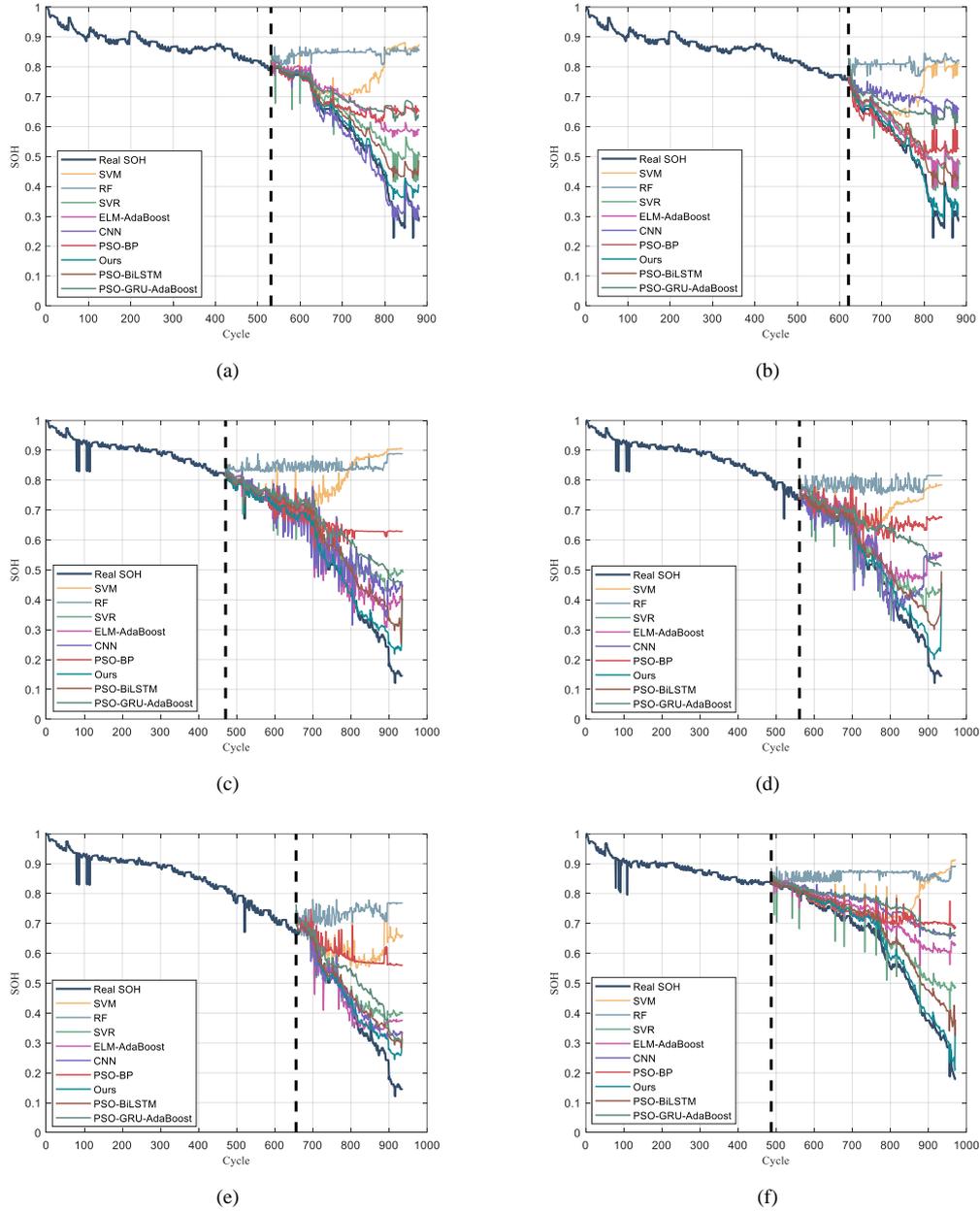

**Fig. 12.** SOH prediction results based on different models for three groups of batteries. (a) CS2-35, Training Data=60%; (b) CS2-35, Training Data=70%; (c) CS2-36, Training Data=50%; (d) CS2-36, Training Data=60%; (e) CS2-36, Training Data=70%; (f) CS2-37, Training Data=50%

In this paper, MAE, RMSE, MSE and MAPE are used as evaluation metrics and Table 5 shows the prediction errors of batteries CS2-35, CS2-36 using different models with different training data. The errors of batteries CS2-37 and CS2-38 are shown in Table 6. As demonstrated in Table 5, our proposed model outperforms the rest of the models in terms of MSE, RMSE, MAPE, and MAE, and the error decreases with the increase of training data, which matches with natural cognition and normal logic. According to the results presented in Table 5, for the battery CS2-35, our proposed method improves by an order of magnitude compared to the traditional SVM, RF and CNN. Compared to PSO-BiLSTM, our model improves the MSE by 60% and 74% when using 50% and 60% of the training data, respectively. It proves that the AdaBoost algorithm can effectively improve the accuracy of the model. In addition, when the training data account for 70%, our models are all one order of magnitude better than other models, indicating that the model proposed in this paper is better at handling longer data. Especially, we present the errors in the predictions of CS2-36 in the form of radar plots to visualize the results, as shown in Fig. 13. It is worth noting that we have uniformly decreased the value of MAPE by three orders of magnitude in order to maintain the consistency of the order of magnitude size. As shown in Table 6, our model performs well ahead of all batteries in CS2-37, predicting results with an order of magnitude higher MSE than the other algorithms. For battery CS2-38, the ability of PSO-BiLSTM to deal with fewer training data after integrating the AdaBoost algorithm becomes weaker. The possible reason is that the model cannot handle the fluctuation at the 746th cycle when there is less training data, which in turn leads to a decrease in model performance. In addition, from Tables 5 and 6, we can see that GRU, as a variant of LSTM, even though it is structurally more streamlined than the LSTM cell, its constituent PSO-GRU-AdaBoost model performs less well than our proposed method on all the four cells, i.e., the model composed of GRU with AdaBoost algorithm is weaker than BiLSTM-AdaBoost when the same parameter optimization strategy is used.

Table 5 SOH prediction results for cell CS2-35 and CS2-36 for different training data using various models.

| Cell | Training Data | Model | MSE | RMSE | MAPE(%) | MAE |
| --- | --- | --- | --- | --- | --- | --- |
| CS2-35 | 50% | SVM | 0.090342 | 0.30057 | 51.602 | 0.20406 |
| | | RF | 0.090052 | 0.3000 | 54.8174 | 0.23864 |
| | | SVR | 0.0087027 | 0.093288 | 16.169 | 0.064696 |
| | | ELM-Ada | 0.0095922 | 0.09794 | 17.7566 | 0.075698 |
| | | CNN | 0.014411 | 0.12004 | 22.1335 | 0.096521 |
| | | PSO-BP | 0.087578 | 0.29594 | 53.617 | 0.23082 |
| | | Ours | **0.0013335** | **0.036517** | **6.5128** | **0.027786** |
| | | PSO-BiLSTM | 0.0033708 | 0.058058 | 9.5382 | 0.035668 |
| | | PSO-GRU-Ada | 0.047534 | 0.21802 | 38.3488 | 0.15666 |
| | 60% | SVM | 0.082042 | 0.28643 | 52.7907 | 0.19708 |
| | | RF | 0.10028 | 0.31601 | 62.9797 | 0.26286 |
| | | SVR | 0.0097028 | 0.098503 | 18.741 | 0.073451 |
| | | ELM-Ada | 0.034225 | 0.185 | 35.1189 | 0.13565 |
| | | CNN | 0.00068802 | 0.02623 | 4.5095 | 0.02108 |
| | | PSO-BP | 0.029382 | 0.17141 | 31.0351 | 0.11325 |
| | | Ours | **0.0012182** | **0.034903** | **6.2125** | **0.022761** |
| | | PSO-BiLSTM | 0.0047506 | 0.068925 | 13.0116 | 0.04981 |
| | | PSO-GRU-Ada | 0.031481 | 0.17743 | 33.7122 | 0.13101 |
| | 70% | SVM | 0.07988 | 0.28265 | 57.3129 | 0.20248 |
| | | RF | 0.1113 | 0.33361 | 74.5217 | 0.29827 |
| | | SVR | 0.0087222 | 0.093393 | 19.619 | 0.072974 |
| | | ELM-Ada | 0.015627 | 0.12501 | 26.9915 | 0.10126 |
| | | CNN | 0.049048 | 0.22147 | 48.4252 | 0.18706 |
| | | PSO-BP | 0.013911 | 0.11794 | 22.9085 | 0.07837 |
| | | Ours | **0.00035571** | **0.01886** | **3.7578** | **0.015587** |
| | | PSO-BiLSTM | 0.0059797 | 0.077329 | 17.1682 | 0.068298 |
| | | PSO-GRU-Ada | 0.037656 | 0.19405 | 42.0747 | 0.16028 |
| CS2-36 | 50% | SVM | 0.12779 | 0.35747 | 89.5102 | 0.24352 |

|  |  | RF | 0.13314 | 0.36488 | 95.5557 | 0.29625 |
|  |  | SVR | 0.016813 | 0.12967 | 33.446 | 0.090464 |
|  |  | ELM-Ada | 0.0066437 | 0.081509 | 19.1698 | 0.047406 |
|  |  | CNN | 0.013743 | 0.11723 | 29.7298 | 0.0821 |
|  |  | PSO-BP | 0.042718 | 0.20668 | 52.3148 | 0.13844 |
|  |  | Ours | **0.0008377** | **0.028943** | **6.615** | **0.014941** |
|  |  | PSO-BiLSTM | 0.0058952 | 0.07678 | 20.0641 | 0.061334 |
|  |  | PSO-GRU-Ada | 0.021538 | 0.14676 | 37.8687 | 0.11204 |
|  | 60% | SVM | 0.091455 | 0.30241 | 84.2003 | 0.21661 |
|  |  | RF | 0.12402 | 0.35217 | 100.7719 | 0.29037 |
|  |  | SVR | 0.012268 | 0.11076 | 31.029 | 0.079081 |
|  |  | ELM-Ada | 0.011087 | 0.10529 | 29.0043 | 0.07147 |
|  |  | CNN | 0.021027 | 0.14501 | 38.7616 | 0.090571 |
|  |  | PSO-BP | 0.061968 | 0.24893 | 69.9018 | 0.18291 |
|  |  | Ours | **0.0014182** | **0.037659** | **9.8772** | **0.02612** |
|  |  | PSO-BiLSTM | 0.063444 | 0.079652 | 22.1393 | 0.056998 |
|  |  | PSO-GRU-Ada | 0.039513 | 0.19878 | 55.458 | 0.15552 |
|  | 70% | SVM | 0.066539 | 0.25795 | 82.4638 | 0.20156 |
|  |  | RF | 0.1327 | 0.36428 | 117.0457 | 0.31442 |
|  |  | SVR | 0.01 | 0.11027 | 34.892 | 0.08325 |
|  |  | ELM-Ada | 0.0058836 | 0.076705 | 22.5987 | 0.052559 |
|  |  | CNN | 0.0058221 | 0.076303 | 23.5335 | 0.055452 |
|  |  | PSO-BP | 0.049901 | 0.22338 | 71.811 | 0.18271 |
|  |  | Ours | **0.0022306** | **0.04723** | **14.3986** | **0.031373** |
|  |  | PSO-BiLSTM | 0.0053819 | 0.073361 | 23.1696 | 0.055009 |
|  |  | PSO-GRU-Ada | 0.012963 | 0.11385 | 35.4255 | 0.099336 |

Table 6 SOH prediction results for cell CS2-37 and CS2-38 for different training data using various models.

| Cell | Training Data | Model | MSE | RMSE | MAPE(%) | MAE |
| --- | --- | --- | --- | --- | --- | --- |
| CS2-37 | 50% | SVM | 0.069022 | 0.26272 | 46.0309 | 0.16157 |
|  |  | RF | 0.088904 | 0.29817 | 57.1878 | 0.23677 |
|  |  | SVR | 0.0086063 | 0.09277 | 17.319 | 0.065716 |
|  |  | ELM-Ada | 0.013527 | 0.1165 | 21.8457 | 0.082903 |
|  |  | CNN | 0.034068 | 0.18458 | 34.9158 | 0.13829 |
|  |  | PSO-BP | 0.036251 | 0.1904 | 34.0037 | 0.12351 |
|  |  | Ours | **0.00060962** | **0.024691** | **4.3181** | **0.021238** |
|  |  | PSO-BiLSTM | 0.004818 | 0.069412 | 13.2969 | 0.056254 |
|  |  | PSO-GRU-Ada | 0.03628 | 0.19047 | 35.8688 | 0.14521 |
|  | 60% | SVM | 0.081792 | 0.28599 | 55.4354 | 0.1913 |
|  |  | RF | 0.094764 | 0.30784 | 64.3164 | 0.25371 |
|  |  | SVR | 0.0091738 | 0.09578 | 19.565 | 0.071464 |
|  |  | ELM-Ada | 0.0204 | 0.14267 | 29.2931 | 0.10803 |
| | | CNN | 0.010579 | 0.10286 | 20.5442 | 0.076221 |
| | | PSO-BP | 0.025039 | 0.15824 | 29.8879 | 0.099696 |
| | | Ours | **0.00071771** | **0.02679** | **5.1103** | **0.019246** |
| | | PSO-BiLSTM | 0.005058 | 0.07112 | 14.1769 | 0.052564 |
| | | PSO-GRU-Ada | 0.0082911 | 0.091055 | 18.4126 | 0.068142 |
| | 70% | SVM | 0.064952 | 0.25486 | 54.4168 | 0.17656 |
| | | RF | 0.087932 | 0.29653 | 67.9121 | 0.24459 |
| | | SVR | 0.0078584 | 0.088648 | 19.98 | 0.068288 |
| | | ELM-Ada | 0.00702 | 0.083832 | 18.9913 | 0.065154 |
| | | CNN | 0.010019 | 0.1001 | 22.4035 | 0.075688 |
| | | PSO-BP | 0.040018 | 0.20004 | 43.524 | 0.14468 |
| | | Ours | **0.00028751** | **0.016956** | **3.5952** | **0.012219** |
| | | PSO-BiLSTM | 0.002078 | 0.045585 | 9.5826 | 0.031062 |
| | | PSO-GRU-Ada | 0.0043129 | 0.065672 | 15.0436 | 0.054791 |
| CS2-38 | 50% | SVM | 0.027384 | 0.16548 | 27.287 | 0.10352 |
| | | RF | 0.086416 | 0.29397 | 52.7705 | 0.22634 |
| | | SVR | 0.0073786 | 0.085899 | 15.024 | 0.062134 |
| | | ELM-Ada | 0.011 | 0.10489 | 18.9261 | 0.080816 |
| | | CNN | 0.040251 | 0.20063 | 36.3277 | 0.1614 |
| | | PSO-BP | 0.041368 | 0.20339 | 34.0341 | 0.13187 |
| | | Ours | 0.002088 | 0.045694 | 8.1713 | 0.038209 |
| | | PSO-BiLSTM | **0.0013887** | **0.037265** | **6.697** | **0.030627** |
| | | PSO-GRU-Ada | 0.044892 | 0.21188 | 37.5485 | 0.15875 |
| | 60% | SVM | 0.02359 | 0.15359 | 26.9656 | 0.097123 |
| | | RF | 0.090521 | 0.30087 | 58.0707 | 0.2359 |
| | | SVR | 0.0067318 | 0.082048 | 15.47 | 0.061273 |
| | | ELM-Ada | 0.011716 | 0.10824 | 21.1155 | 0.086571 |
| | | CNN | 0.00094824 | 0.030794 | 5.2358 | 0.023124 |
| | | PSO-BP | 0.022514 | 0.15005 | 27.0186 | 0.10002 |
| | | Ours | 0.00046465 | 0.021556 | 3.7982 | 0.01398 |
| | | PSO-BiLSTM | **0.0000447** | **0.0066866** | **0.99062** | **0.004664** |
| | | PSO-GRU-Ada | 0.0070419 | 0.083916 | 15.698 | 0.060806 |
| | 70% | SVM | 0.019615 | 0.14005 | 27.326 | 0.095016 |
| | | RF | 0.098366 | 0.31363 | 66.7937 | 0.25564 |
| | | SVR | 0.005063 | 0.071154 | 14.4916 | 0.054037 |
| | | ELM-Ada | 0.0027278 | 0.052228 | 10.9103 | 0.042542 |
| | | CNN | 0.040372 | 0.20093 | 43.3366 | 0.17136 |
| | | PSO-BP | 0.024987 | 0.15807 | 33.2714 | 0.12525 |
| | | Ours | **0.00015654** | **0.012511** | **2.3775** | **0.0095594** |
| | | PSO-BiLSTM | 0.00023344 | 0.015279 | 2.4292 | 0.010005 |
| | | PSO-GRU-Ada | 0.0070139 | 0.083749 | 18.1197 | 0.071113 |

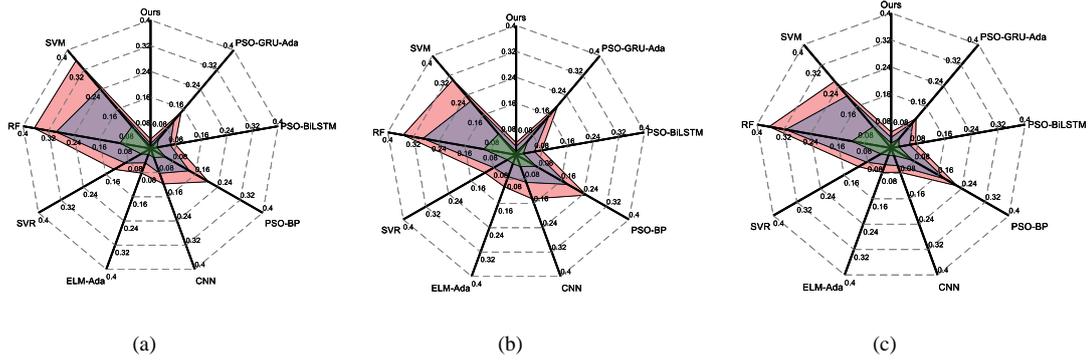

**Fig. 13.** RMSE error for CS2-36. (a) Training Data=50%; (b) Training Data=60%; (c) Training Data=70%. In (a), (b) and (c), red, blue and green color represent RMSE, MAE and MAPE, respectively.

Based on the experimental results, the factors affecting the accuracy of the SOH prediction model proposed in this paper are summarized as the following three.

(1) Structure and parameters of the model. This paper combines the advantages of BiLSTM and AdaBoost algorithm. Compared with LSTM, BiLSTM can capture the features contained in the sequences more comprehensively, while AdaBoost has the ability to prevent the model from overfitting, and at the same time, it can improve the generalization of the model. For the model parameters, this paper chooses the PSO algorithm to obtain the learning rate and the number of neurons of BiLSTM rather than relying on experience to set them, so as to improve the prediction accuracy.

(2) Setting of training data. From Section 4.2, it can be seen that when the training data is more, the prediction effect of the model is more excellent. However, the AdaBoost algorithm instead leads to a slight degradation of the model's performance when there is less training data.

(3) Effect of aging inflection point on the model. As demonstrated in battery CS2-35, if the starting point of the model's prediction is located near the battery's aging inflection point, the rapidly decreasing data can cause difficulties in the model's accurate prediction. For this problem we can develop a strategy: when using less data, the PSO-BiLSTM model is used for prediction; when more training data is available the AdaBoost algorithm is fused to obtain good prediction results.

## 5. Conclusions

In this paper, a fusion model incorporating PSO-BiLSTM and AdaBoost algorithms is designed for predicting the SOH of LiBs. It combines the structure and advantages of BiLSTM and AdaBoost to show high accuracy on the prediction task of SOH. The test results based on public datasets reflect the advantages of the model proposed in this paper: higher accuracy, better robustness and generalization ability. For batteries CS2-35, CS2-36, and CS2-37, the RMSE of the present method is improved by at least 33%, 35%, and 61%, respectively, and up to 94%, 92%, and 94%, respectively, and exhibits the best performance on battery CS2-37. Future work can be considered as follows: (1) Optimize the model structure and adopt more advanced optimization algorithms to reduce the parameter search time and time cost (2) Construct SOH prediction of LiBs based on the actual state of use, such as unmanned aerial vehicles, electric vehicles, and so on.